\documentclass[12pt]{article}

\usepackage{epsfig}
\usepackage{graphicx}
\usepackage{psfrag}
\usepackage{amssymb}
\usepackage{amsmath}

\newcommand{\rf}[1]{(\ref{#1})}
\newcommand{\beq}{\begin{equation}}
\newcommand{\beql}[1]{\beq\label{#1}}
\newcommand{\eeq}{\end{equation}}
\newcommand{\bea}{\begin{eqnarray}}
\newcommand{\eea}{\end{eqnarray}}

\newcommand{\e}{\mbox{e}}
\renewcommand{\d}{\mbox{d}}
\newcommand{\g}{\gamma}

\newcommand{\lam}{\lambda}
\newcommand{\La}{\Lambda}
\renewcommand{\b}{\beta}
\renewcommand{\a}{\alpha}
\newcommand{\n}{\nu}

\newcommand{\ep}{\varepsilon}

\newcommand{\om}{\omega}

\newcommand{\del}{\delta}
\newcommand{\Del}{\Delta}

\newcommand{\kp}{\kappa}

\newcommand{\oh}{\frac{1}{2}}

\newcommand{\ra}{\rangle}
\newcommand{\la}{\langle}

\newcommand{\prt}{\partial}
\newcommand{\mi}{\!-\!}

\newcommand{\cD}{{\cal D}}

\newcommand{\cN}{{\cal N}}

\newcommand{\cO}{{\cal O}}

\newcommand{\tG}{{\tilde{G}}}
\newcommand{\tL}{{\tilde{\La}}}
\newcommand{\tV}{{\tilde{V}}}

\newcommand{\tx}{{\tilde{x}}}

\newcommand{\ty}{{\tilde{y}}}

\newcommand{\tc}{{\tilde{c}}}

\newcommand{\hG}{{\hat{G}}}

\newcommand{\hH}{{\hat{H}}}

\begin{document}

\vspace{-36pt}

\begin{center}

{ \large \bf Quantum Gravity via Causal Dynamical Triangulations}

\vspace{36pt}

{\sl J. Ambj\o rn}$\,^{a,c}$,
{\sl A. G\"{o}rlich}$\,^{a,b}$,
{\sl J. Jurkiewicz}$\,^{b}$ and
{\sl R. Loll}$\,^{c}$

\vspace{24pt}

{\footnotesize

$^a$~The Niels Bohr Institute, Copenhagen University\\
Blegdamsvej 17, DK-2100 Copenhagen \O , Denmark.\\
{email: ambjorn@nbi.dk}\\

\vspace{6pt}

$^b$~Institute of Physics, Jagellonian University,\\
Reymonta 4, PL 30-059 Krakow, Poland.\\
{ email: atg@th.if.uj.edu.pl, jurkiewicz@th.if.uj.edu.pl}\\

\vspace{6pt}

$^c$~Radboud University Nijmegen\\
Institute for Mathematics, Astrophysics and Particle Physics (IMAPP)\\ 
 Heyendaalseweg 135,
6525 AJ Nijmegen, The Netherlands. \\
{email: r.loll@science.ru.nl, j.ambjorn@science.ru.nl}

}
\vspace{24pt}
\end{center}

\begin{center}
{\bf Abstract}
\end{center}


\noindent ``Causal Dynamical Triangulations'' (CDT) represent a 
lattice regularization of the sum over spacetime histories,
providing us with a non-perturbative formulation of quantum gravity. 
The ultraviolet fixed points of the lattice theory can be used to 
define a continuum quantum field theory, potentially making 
contact with quantum gravity defined via asymptotic safety. 
We describe the formalism of CDT, its phase diagram, 
and the {\it quantum geometries} emerging from it.
We also argue that the formalism should be able to describe a more
general class of quantum-gravitational models of Ho\v rava-Lifshitz type.

\newpage

\section{Introduction}
\label{intro}

At this stage, there is no certainty how to best reconcile the 
classical theory of relativity with quantum mechanics. 
Applying the well-tested methods of quantization to gravity -- defined by
the Einstein-Hilbert action -- and quantizing the fluctuations
around a classical solution to Einstein's equations leads to a 
non-renormalizable theory. 
This happens because in four spacetime dimensions the mass dimension of 
the gravitational coupling constant $G$ (in units where $\hbar$ and $c$
are 1) is $-2$, whereas it should be larger than or equal to 0 for the theory to be 
renormalizable perturbatively. One would therefore expect the perturbative 
effective quantum field theory description to break down at energies $E$
satisfying $G E^2 \gtrsim 1$. 

There are of course well-known examples where the non-renormalizability 
of a quantum field theory in the ultraviolet (UV) was eventually resolved  
by introducing new degrees of freedom, missed initially because they were not directly 
observable at low energies. 
The electroweak theory is an example where perturbative renormalizability
was ``regained" in this way. The theory was described first by a four-fermion interaction
with an associated Fermi coupling $G_F$ of mass dimension $-2$, just like
the Newton constant $G$ in gravity. As a result, its perturbation theory breaks down 
at energies with $G_F E^2 \gtrsim 1$. However, it turns out that
for energies above $1/\sqrt{G_F}\approx M_W$, the mass 
of the $W$-particle, the four-fermion theory has to be replaced 
by the $SU(2)$-gauge theory of the weak interactions, which contains new excitations,
the $W$- and $Z$-bosons. The new electroweak theory 
{\it is} a renormalizable quantum field theory. 

Similarly, in the 1960s
the low-energy scattering of pions was described by a non-linear 
sigma model, another non-renormalizable quantum field theory 
whose coupling constant, the pion decay constant $F_\pi$-squared, 
has mass dimension $-2$. However, high-energy scattering at
energies beyond $1/F_\pi$ is no longer described well by the non-linear 
sigma model, because it starts probing the intrinsic structure of the
pions. A correct description has to incorporate appropriate new degrees of freedom,
the quarks and gluons, and the corresponding quantum theory --
quantum chromodynamics -- is perfectly renormalizable. 

There is no obvious reason which prevents us from writing down a perturbative
(and non-renormalizable) expansion for gravity around some classical background 
geometry, say,
flat Minkowski spacetime, if we are interested in an effective quantum 
field-theoretic description whose range of applicability does not extend
beyond energies with $G  E^2 \approx 1$. In view of the examples cited above,
it is then tempting to conjecture that the apparent non-renormalizability
of gravity could be resolved by the appearance of new degrees of freedom
at higher energies, rendering the theory renormalizable after all. 

A solution of this kind may be in the form of a superstring theory in a 
higher-dimensional spacetime, where the gravitational excitations are 
intertwined with infinitely many new degrees of freedom in such a way as to cure
the UV problem. Although string theory cannot be ruled out as the correct answer,
the world picture it provides has yet to be verified. In particular, 
supersymmetry -- predicted by string theory -- has not yet
been observed at the Large Hadron Collider. Of course, even if no evidence
of supersymmetry is found at this or future colliders, it may still be present
at even higher energies. In this sense, the absence of observational
evidence for supersymmetry does not disprove
superstring theory as such, although it makes it less compelling as a resolution
of the problem of unifying gravity and quantum theory. 

There are other potential resolutions to the problem of finding a suitable
``ultraviolet completion" of perturbative quantum gravity, which are not based
on fundamental, string-like excitations and do not obviously require the existence of
supersymmetry or extra dimensions. These are so-called
non-perturbative approaches, whose starting point typically consists of a set of dynamical 
degrees of freedom closely modeled on those of classical gravity (``curved geometry" in one way or
other), together with a non-perturbative prescription for quantization. A concrete
example, that of {\it Causal Dynamical Triangulations}, will be described in some detail
below. Its geometric degrees of freedom, in presence of a UV cut-off, are given in
terms of triangulated, piecewise flat spacetimes with discrete curvature assignments.
Its non-perturbative quantization follows that of a
standard lattice field theory, albeit with a dynamical rather than a fixed lattice. 

An obvious charm of such a purely quantum field-theoretic ansatz lies in its minimalism, 
and the absence -- to a 
large degree -- of free parameters and other ``tunable" ingredients. 
On the other hand, a key
difficulty of this type of approach is to demonstrate that it is related to classical
gravity in a suitable limit, something that is not at all obvious once one has moved beyond 
linearized quantum fields on a fixed background spacetime.
One also needs to spell out what it means for the non-perturbative theory to
{\it exist}, which likewise is non-trivial in a background-free description where 
``observables" are hard to come by. 

In parallel with advances in string theory, 
also research in the wider area of 
non-perturbative quantum gravity has seen a steady rise in interest in recent 
decades. On the one hand, this was due to the rejuvenation of canonical quantum gravity in the
form of {\it loop quantum gravity} from the late 1980s onwards.\footnote{Curiously, this ansatz also
postulates the fundamental character of certain one-dimensional ``closed-string" (a.k.a. ``loop")
excitations in the quantum theory.} At
about the same time, the covariant {\it gravitational path integral}
was given a new, non-perturbative lease of life in terms of ``dynamical triangulations".
Motivated originally by the search for a non-perturbative dynamics of curved, two-dimensional 
worldsheets in (bosonic) string theory, this dynamical lattice formulation provides a powerful
computational tool for evaluating gravitational path integrals quantitatively: analytically in two, and 
numerically in higher dimensions. --
The focus of the present article will be on this latter development, arguably the 
conceptually most straightforward and methodologically minimalist extension of
the standard perturbative and covariant quantum field-theoretic formulation of gravity. 
We will explain how it may lead to the
construction of a viable theory of quantum gravity, valid on all scales, without running into
contradictions vis-\`a-vis the perturbative non-renormalizability of the theory.

In the late 1970s, Weinberg outlined a scenario, coined {\it asymptotic safety} \cite{weinberg}, 
for how quantum field theories which are not power-counting 
renormalizable around a trivial Gaussian fixed point could under certain, general
conditions still make sense, just like ordinary renormalizable theories.
In particular, an asymptotically safe theory is characterized by only a 
{\it finite} number of coupling constants, whose values will be determined by 
comparison with experiment or observation. 
The asymptotic freedom scenario is naturally described in the language of
quantum field theory and the renormalization group. It is characterized by
the presence of an ultraviolet fixed point in the infinite-dimensional 
coupling constant space of a theory, with the property that in the fixed point's
neighbourhood the dimension of the subspace of attraction is infinite-dimensional,
with finite co-dimension. This co-dimension coincides with the number
of free parameters of the theory that need to be fixed by experiment.  
Such a UV fixed point therefore attains a similar status to that of
the Gaussian fixed point of a renormalizable theory.
The snag is that the tools of the perturbative theory are usually not sufficient
to find such ultraviolet fixed points -- if they exist for a given theory -- and to study 
their neighbourhoods.
 
To illustrate the implications of the presence of such a fixed point (in a somewhat simplistic fashion),
let us introduce the dimensionless coupling 
\beq\label{gcoupl}
\tG (E):= G E^2. 
\eeq
A fixed point in this context always 
refers to the behaviour under a change of scale $E$ of a dimensionless, energy-dependent function 
like $\tG(E)$. The dimensionful quantity $G$ in \rf{gcoupl} can at this stage still be thought of as
a (classical, low-energy) coupling {\it constant} of mass dimension $-2$.
Let the behaviour of $\tG(E)$ be dictated by a beta function $\b(\tG)$ according to
\beq\label{h1}  
 E \frac{\d \tG}{\d E} = \b(\tG),~~~~{\rm with}~~\b(\tG) = 2 \tG - 2 \om \tG^2, 
\eeq
for some real parameter $\om$. It is immediately clear that for $\om \not= 0$, $G=const$ is
no longer a solution to \rf{h1}. For consistency, $G$ has to acquire a non-trivial $E$-dependence 
and therefore becomes a {\it function} $G(E)=\tG(E)/E^2$.
In \rf{h1} we have chosen the simplest non-trivial beta function such that (i) in the limit of low energy,
$E \to 0$, $G(E)$ goes to a constant (which we will continue to call $G$), and 
(ii)  for $E \to \infty$, $\tG(E)$ goes to a non-trivial 
UV fixed point. Explicitly, the solution to the differential equation in \rf{h1} can be stated as
\beq\label{h2}
G(E) =  \frac{G}{1+\om G E^2},
\eeq 
from which we can read off the location of the UV fixed point 
at $\tG = 1/\om$, the non-trivial zero of the beta function. 
An important feature of this solution is that 
the coupling constant $G(E)$ goes to zero at the UV fixed
point. 

\begin{figure}
\psfrag{B}{{\bf{\LARGE $\b(g)$}}}
\centerline{\scalebox{0.4}{\rotatebox{0}{\includegraphics{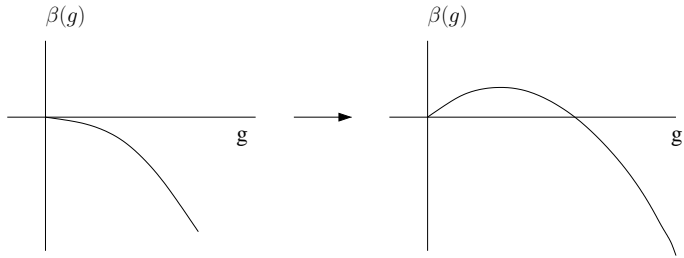}}}}
\caption{Changing an asymptotically free theory to an asymptotically safe one by
increasing its dimension from $d$ to $d+\ep$ results in a shift of
its ultraviolet fixed point to a value $g>0$.}
\label{figh1}
\end{figure}

In case the above example should appear somewhat ad hoc, it can be understood as arising 
from a more general construction, which starts from an asymptotically {\it free} theory in
$d$ dimensions. Fig.\ \ref{figh1} (left) illustrates the corresponding (negative) beta function
of the coupling $g$, together with a Gaussian UV fixed point at $g=0$. 
If this theory is  ``lifted'' to $d+\ep$ dimensions -- assuming that such a perturbation in the
dimension is well defined, at least for small $\ep >0$ -- its beta function will change according to
\beq\label{h3}
\b(g) \to \rho(\ep) g + \b(g),
\eeq
where $\rho(\ep)$ is the (positive) amount by which the mass dimension of the coupling $g$ 
{\it de}creases
as a result of the dimensional increase by $\ep$. 
(Our previous example, whose beta function was defined in relation \rf{h1}, corresponds to $\rho=2$.)
Note that the Gaussian UV fixed point of the original theory has become a 
non-trivial UV fixed point away from zero in the higher-dimensional theory, while $g=0$ has been
turned into an infrared fixed point, as illustrated by Fig.\ \ref{figh1}.

The theories we have discussed so far -- four-Fermi theory, non-linear sigma model 
and Einstein gravity -- display a similar behaviour in the sense that they are
asymptotically free, renormalizable theories in spacetime dimension $d=2$.
Trying to make sense of them beyond dimension 2 by way of a
$2+\ep$-expansion, one encounters the situation depicted in Fig.\ \ref{figh1}.
Of course, one may formally set $\ep =2$ in such an expansion, as would be needed 
to reach the dimension $d=4$ of physical spacetime, but the validity of the perturbative
expansion for such large values of $\ep$ would need to be established to take 
the results seriously, and a priori appears perhaps rather doubtful.

Non-trivial UV-complete extensions to $d=4$ of the four-Fermi interaction
or the non-linear sigma model are not known and presumably do not exist.
As mentioned above, we should rather think of them as effective 
theories, which happen to describe certain low-energy properties of 
more fundamental theories with more and different fundamental excitations.
Still, it is difficult to draw any conclusions from this for general relativity, the theory we are
interested in, which is after all very different physically: exactly the
degrees of freedom that are fixed in all other theories, those of spacetime itself,
become dynamical in gravity. Much work has gone into trying to
show that four-dimensional gravity possesses an ultraviolet fixed point with the requisite
properties, either in terms of the $2+\ep$-expansion \cite{kawai} or by
using general renormalization group techniques \cite{rg}.
 
In what follows, we will not be concerned with the details of these efforts, 
but with the question of how the hypothesis of asymptotically safe gravity may be 
tested independently and non-perturbatively by using standard field-theoretic tools
and by formulating quantum gravity via a lattice regularization.

\section{A lattice theory for gravity}

A number of issues have to be addressed when representing gravity on a lattice.
Is it possible in principle to construct a well-defined lattice regularization of
gravity\index{Lattice Gravity} with a UV lattice cut-off, 
which can be removed in a controlled way
to obtain a continuum limit (whatever this may turn out to be)? 
The answer is {\it yes}. More precisely,
the issue is not so much how to represent gravity on a lattice, but
how to represent a theory as a lattice theory whose standard continuum
formulation in terms of local fields is diffeomorphism-invariant, a vast gauge 
invariance closely related to the differentiable structure of the underlying
manifold and its description in terms of local coordinate charts.

For the geometric degrees of freedom of the gravitational theory this 
can be done by viewing the lattice 
itself as representing directly a (piecewise linear) geometry. The key point is
that such a geometry can be described uniquely without ever introducing
coordinates, thus circumventing the associated redundancy of having to choose
any particular set of coordinates. A convenient choice is to use lattices which are
triangulations, in the sense of consisting of $d$-simplices, 
triangular building blocks which are $d$-dimensional generalizations of
flat triangles ($=$2-simplices). Assuming the interior of a $d$-simplex to be
flat, its geometry is uniquely specified by giving the lengths of its 
$d(d+1)/2$ one-dimensional edges or links. Together with the information of how
the simplices are ``glued together" (that is, how $(d-1)$-dimensional boundary 
simplices are identified pairwise) to form a triangulated manifold, this
suffices to compute all geometric information, including distances, geodesics, 
volumes etc. without using coordinates. 
Important for our path integral representation, Regge observed
that the curvature of such a piecewise linear geometry is in a natural way located 
on its $(d-2)$-dimensional subsimplices (the ``hinges"). 
By the same token, the scalar curvature term of the Einstein action of 
such a geometry is given by the sum over all hinges of the deficit angle
around each hinge, multiplied by the hinge's volume \cite{regge}.

In our construction of a theory of quantum gravity, the lattice-regularized path integral  
over geometries thus becomes the sum over such triangulations, with weight
depending on the Regge implementation of the Einstein action. 
Precisely which class of triangulations should we sum over in the path integral?
When applying Regge calculus to classical gravity 
one uses a fixed lattice, in the sense of leaving the connectivity of its constituent simplicial
building blocks unaltered. This still allows the curvature of the triangulation to be changed -- for
example, to optimally approximate that of a given smooth geometry -- by changing the
lengths of its one-dimensional edges. 

When using the piecewise linear geometries 
in a path integral, the task is different. Firstly, we do not expect the 
individual path integral configurations to be smooth, but only continuous,
in the same way as the paths in the path integral of a quantum-mechanical particle are 
continuous but in general non-smooth\footnote{in fact, with unit probability
they are nowhere differentiable}. Similarly, the piecewise linear geometries are
a subset of all continuous spacetime geometries. Note that we can even restrict 
ourselves to a subset of piecewise linear geometries as long as it
is suitably dense in the set of all geometries. More precisely, when the 
lattice spacing goes to zero, we require the expectation values of observables, again
suitably defined on the piecewise linear geometries, to converge 
to the value they would take in the continuum quantum field theory 
(which we assume exists). In contrast with the aim of the classical theory,
we are therefore not trying to approximate 
any {\it particular} geometry by our lattice geometries, but to span the 
whole set of geometries. 

In this context a specific subset of piecewise 
linear geometries has proved to be very useful, namely, the triangulations  
whose edges have all the same length, $a$, say. One can 
characterize this set of geometries as being constructed from
gluing together {\it equilateral} simplicial building blocks
in all possible ways, compatible with certain 
constraints (typically, a fixed topology and fixed boundary components). 
Consequently, the variation in geometry (the way in which the geometric 
degrees of freedom are encoded) is linked to the mutual connectivity of the building blocks
created by the gluing and not to variations in the link lengths, giving rise to the 
name {\it Dynamical Triangulations} \index{Dynamical Triangulations}(DT) 
\cite{adf,david,kazakov}. 
From a path-integral perspective this approach has the advantage 
that distinct triangulations correspond to physically
distinct geometries. Summing 
over this DT ensemble of geometries may therefore lead directly 
to the correct continuum measure in the limit that the UV cut-off is taken to zero, $a \to 0$. 
By contrast, treating the triangulations classically \`a la
Regge, with fixed lattice connectivity and variable link lengths,
still contains redundancies, in the sense that many different 
lattice configurations can correspond to the same physical geometry (see \cite{diffeos}
and references therein). 
For illustration, consider a rectangle in the two-dimensional plane and 
triangulate its interior. 
Clearly, the interior vertices can be moved around locally in the plane 
without changing the flat geometry of the rectangle. However, since all of these
are different as Regge triangulations, this leads to a severe overcounting
in the path integral of quantum Regge calculus, for which there is currently no known fix.
 
Most importantly, the viability of the DT lattice regularization has already been demonstrated 
in a non-trivial case, that of gravity (coupled to matter) in two dimensions. 
As mentioned above, two-dimensional 
gravity is a renormalizable quantum field theory and various observables
can be calculated analytically \cite{KPZ}. 
The dynamically triangulated two-dimensional lattice
theory can also be solved, a number of observables can be 
calculated analytically and its continuum limit, taking the lattice spacing $a\to 0$,
can be taken \cite{DT}. Remarkably, results from the two different 
calculations can be compared
and are found to agree. 
We conclude that {\it it is possible to provide a viable lattice regularization
of a diffeomorphism-invariant quantum theory of geometries.} 

One may object that this two-dimensional theory has little to do with true
gravity in four spacetime dimensions; to start with, it has no propagating
gravitons. However, we would like to argue that it is much more a theory of 
fluctuating geometries than one would ever expect of the four-dimensional
theory. Because there is no Einstein-Hilbert action in two dimensions 
(it is topological), each configuration contributes in the path integral with 
the same weight, which is a maximally quantum situation. 
This is borne out by the analytic solutions of this model, which show 
the two-dimensional geometries as wildly quantum-fluctuating. 
Nevertheless the lattice theory has no problem in reproducing
the correct diffeomorphism-invariant continuum theory, also known as 
quantum Liouville gravity.

\subsection{Observables}\label{2.1}

How to define what does and does not constitute an ``observable" in quantum gravity, 
and how to construct and evaluate observables in any given formulation are physical
questions of central importance \index{Observables in quantum gravity}. 
What we would like to highlight here is that a 
beautiful aspect of a geometric lattice formulation of quantum gravity 
of the type we are considering is that it forces one 
to address such questions head-on. It is 
not possible to hide behind some ``expansion around flat spacetime'', but one is
forced to think in terms of physical ``rods and clocks", much in the spirit of Einstein's
classical theory.

Let us discuss the basic objects of any quantum field theory, namely,
the correlators of local quantum operators $\cO(x)$. Such correlators
are important ingredients in constructing S-matrix elements, i.e.\ observables
in quantum field theory on a fixed background. Also in conventional lattice 
theories, correlators play a crucial role in  
showing that a lattice theory has a continuum limit when the 
lattice spacing goes to zero. 

Consider some lattice scalar field theory, and let $\cO(x_n)$ be an operator at lattice 
spacetime coordinate $x_n = n \cdot a$, where $a$ is the lattice spacing 
and $n$ the integer-valued
lattice coordinate. In general, we expect the correlator to fall off
exponentially,
\beq\label{h4}
-\log \la  \cO(x_n) \cO(x_m) \ra \sim |n-m|/ \xi(g_0) + o(|n-m|),
\eeq
where $g_0$ is the bare lattice coupling and $\xi(g_0)$ the 
correlation length in lattice spacings. 
The standard procedure for a lattice
system is to take the continuum limit at a second-order phase transition point 
$g_0^c$, where the correlation length diverges like 
\beq\label{h5}
\xi(g_0)\propto \frac{1}{|g_0-g_0^c|^\n},~~~~~
a(g_0)\propto |g_0-g_0^c|^\n.
\eeq
Eq.\ \rf{h5} tells us at what rate we should scale the lattice 
spacing to zero in the limit $g_0 \to g_0^c$, in order to find an exponential 
decay in the continuum, when the lattice correlation 
diverges, but the (dimensional) physical length 
$x_n -x_m= (n-m)a$ is kept constant,
\beq\label{h6}
m_{ph} a(g_0) = 1/\xi(g_0),~~~~\e^{-|n-m|/\xi(g_0)} = \e^{-m_{ph}|x_n-x_m|}.
\eeq   
Eq.\ \rf{h6} illustrates the fact that dimensionful observables, 
like the physical mass $m_{ph}$, are defined by the {\it approach} to the critical point, not 
{\it at} the critical point. 
    
The existence of a critical point \index{Critical Point}
and an associated divergent correlation length 
constitute the backbone of the Wilsonian renormalization group approach to
quantum field theory. Since we are appealing to this Wilsonian approach
by asking whether asymptotic safety is realized, it is important to 
understand whether it can be applied to quantum gravity at all. A first step 
in this direction is to understand whether suitable correlators 
\index{Correlator in Quantum Gravity} and a correlation length
can be defined in a diffeomorphism-invariant theory like quantum gravity.
To start with, how can we define the distance
between two points in a path integral where we integrate over
the geometries defining this distance?

In flat $d$-dimensional spacetime, let us rewrite the correlator of  
a scalar field $\phi(x)$, say, in the form
\beq\label{h7}
\la \phi \phi (R) \ra_V \equiv 
\frac{1}{V} \;\frac{1}{s(R)} \int \cD \phi\, e^{-S[\phi]} 
\int\!\! d^dx \!\int \!\! d^dy
\;  \phi(x) \phi(y)   
\;\del(R \mi |x-y|).   
\eeq
As indicated, this expression depends on a chosen distance $R$, but no longer on specific 
points $x$ and $y$, which instead are integrated over. The integrand can be read 
``from right to left" as first averaging 
over all points $y$ at a distance $R$ from some fixed point $x$, normalized
by the volume $s(R)$ of the spherical shell of radius 
$R$, and then averaging 
over all points $x$, normalized
by the total volume $V$ of spacetime. 
We assume translational and 
rotational invariance of the theory and that $V$ is so large that 
we can ignore any boundary effects related to a finite volume.

This definition of a correlator is of course non-local, but unlike
the underlying locally defined correlator has a straightforward diffeomorphism-invariant 
generalization to the case where gravity is dynamical, namely,
\bea
\la \phi \phi (R) \ra_V &\equiv& 
\frac{1}{V}\int \cD [g]\, \int \cD_{[g]} \phi \;\e^{-S[g,\phi]} 
\;\del\Big(V-\!\!\int \!\! d^dx \sqrt{\det g} \Big)\cdot \nonumber \\  
&&  
\int\!\! d^dx \!\int \!\! d^dy\, \;\frac{\sqrt{\det g(x)} \, \sqrt{\det g(y)}}{s_{[g]}(y,R)}
\; \phi(x) \phi(y)  
\;\del(R \mi D_{[g]}(x,y)),  
\label{h8}
\eea
which now includes a functional integration over geometries\footnote{in accordance
with standard notation, $[g]$ denotes an equivalence class of metrics $g$ under
the action of the diffeomorphism group} $[g]$, and
dependences of the action, measures, distances and volumes on $[g]$.
Can the definition \rf{h8} be implemented meaningfully to define correlators in a
quantum gravity theory? The answer is yes, and a two-dimensional example 
can again be used to demonstrate this. Namely, there are analytic predictions for 
the behaviour of the propagators of certain matter theories coupled
to two-dimensional Euclidean gravity \cite{KPZ}, which have been shown
to be reproduced by numerical simulations of the corresponding lattice theory \cite{aa}.
By the way, their behaviour is quite different from that 
of the flat space correlators, another manifestation of the fact that two-dimensional 
gravity is a theory of strong geometric fluctuations.

\subsection{Time-slicing and baby universes}\label{sectimeslice}

An interesting aspect that can be analyzed in detail 
in the solvable two-dimensional quantum theory of fluctuating geometry is that 
of proper time \index{Proper Time}. One usually considers a 
situation where the rotation to Euclidean signature has taken place and 
``proper time" is simply given by ``geodesic distance". 
In this setting, a closed one-dimensional spatial universe of fixed ``time"  is simply 
a loop of length $\ell$. 
In the corresponding quantum theory one can ask for 
the amplitude for a universe of length $\ell_1$ to 
``propagate'' to another one of length $\ell_2$ in proper time 
$t$. More precisely, the outgoing loop of length $\ell_2$ 
is said to have a proper-time (in this case a geodesic) distance $t$
to the incoming loop of length $\ell_1$ if each point on $\ell_2$ has 
geodesic distance $t$ to $\ell_1$. 
(The geodesic distance from a point 
to a set of points is defined as the minimum of the geodesic distances 
from the point to the points in the set.)
 
\begin{figure}[t]
\centerline{\scalebox{0.4}{\rotatebox{0}{\includegraphics{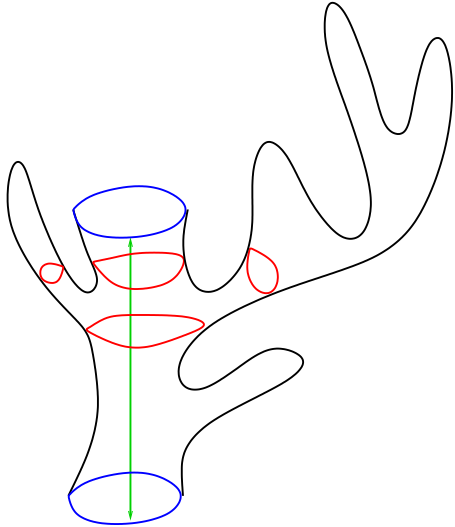}}}}
\caption{Incoming and outgoing boundary loops of length $\ell_1$ and $\ell_2$,
separated by a geodesic distance $t$, and a typical interpolating geometry of
cylinder topology which contributes to the amplitude $G(\ell_1,\ell_2;t)$
in Euclidean
signature. The additional loops drawn onto the interior geometry
consist of points which share the same distance to the incoming loop. 
As indicated by the upper set of three loops, there can be 
many disconnected loops at a given distance to the incoming loop.}
\label{figh2}
\end{figure}
Fig.\ \ref{figh2} shows a typical geometry in 
the path integral contributing to the corresponding amplitude $G(\ell_1,\ell_2;t)$. 
It will often be convenient to work with its Laplace transform,
\beq\label{h9}
G(x,y;t)= \int_0^\infty\int_0^\infty d\ell_1 d\ell_2 \; \e^{-x\ell_1-y\ell_2}
\;G(\ell_1,\ell_2;t).
\eeq
We can view $x$ and $y$ in this expression as {\it boundary cosmological 
constants}, since $x\cdot \ell$ would be the action of a one-dimensional 
``spacetime'' of volume $\ell$ and cosmological constant $x$. 

As shown in \cite{watabiki},
the amplitude $G(x,y;t)$ satisfies the remarkably simple equation
\beq\label{h10}
\frac{\prt G(x,y,t)}{\prt t}= \frac{\prt (W(x) G(x,y,t))}{\prt x},
\eeq
where $W(x)$ is the Hartle-Hawking disk amplitude, which in 
two-dimensional Euclidean gravity is given by \cite{DT}
\beq\label{h11}
W(x)= (x-\oh)\sqrt{x+\sqrt{\La}}.
\eeq
As is clear from Fig.\ \ref{figh2}, space can branch out into 
many disconnected parts (i.e. change its topology) as a function of 
proper time $t$, giving rise to ``baby universes''.
The appearance of baby universes \index{Baby Universes} 
on all scales leads to the two-dimensional 
quantum spacetime being fractal \index{Fractal Spacetime}, 
with Hausdorff dimension \index{Hausdorff Dimension} $d_h =4$ 
\cite{watabiki,aw}.  

Rather amazingly, it is possible to integrate analytically over these baby 
universes, resulting (for each time history) in a spacetime with a
proper-time foliation and no baby universes \cite{ackl}. 
Alternatively, the expression for the loop-loop propagator 
without baby universes can be obtained directly by 
summing over a class of two-dimensional spacetimes which
from the outset lack baby universes,
provided one redefines the coupling constants suitably \cite{al}. 
This latter procedure can be 
implemented also at the regularized level in terms of a set
of ``{\it causal} dynamical triangulations'' 
\index{Causal Dynamical Triangulations} (CDT), to be distinguished
from the larger class of merely ``dynamical triangulations" (DT), which served
as carrier space for the Euclidean gravitational path integral \cite{al}. 

The resulting theory has a well-defined Hamiltonian and 
corresponding unitary proper-time evolution. 
The explicit map between the cosmological constants of DT and CDT turns
out to be non-analytic,
\beq\label{h12}
\tL_{cdt} = \sqrt{\La_{dt}},~~~~~~\tilde{x}_{cdt} = \sqrt{x+\sqrt{\La_{dt}}},
\eeq
where we have denoted the CDT-analogues of the couplings with a subscript and tilde.
Consequently, in CDT both lengths and areas acquire a dimensionality different from
that found in the DT ensemble of spacetimes and in Liouville gravity. When using the 
CDT ensemble, also the
Hausdorff dimension changes from 4 to 2, the canonical
value for ordinary smooth two-dimensional spacetimes.\footnote{A word of warning: 
the coincidence
in Hausdorff dimension does {\it not} allow one to conclude that the quantum geometry
of two-dimensional CDT in any way approximates a smooth classical manifold; in fact,
it does not.}

The CDT loop-loop propagator satisfies the equation 
\beq\label{h13}
\frac{\prt \tG(\tx,\ty,t)}{\prt t}= 
\frac{\prt((\tx^2-\tL_{cdt}) \tG(\tx,\ty,t))}{\prt \tx},
\eeq
and the Hamiltonian governing the (proper-) time evolution is
given by 
\beq\label{h14}
  \tG(\tilde{\ell}_1,\tilde{\ell}_2,t)= 
\la\tilde{\ell}_2|\e^{-t \hH}|\tilde{\ell}_1\ra,~~~~~
\hH= -\tilde{\ell} \frac{d^2}{d\tilde{\ell}^2} +\tL_{cdt} \tilde{\ell},
\eeq
while the CDT Hartle-Hawking wave function 
\index{Hartle-Hawking Wave Function} (which is derived
from the propagator $\tG$ \cite{al}) satisfies
\beq\label{h15}
\hH \tilde{W}_{cdt} (\tilde{\ell})=0. ~~~~~\mbox{ (Wheeler-DeWitt)}
\eeq

\begin{figure}[t]
\centerline{\scalebox{0.5}{\rotatebox{0}{\includegraphics{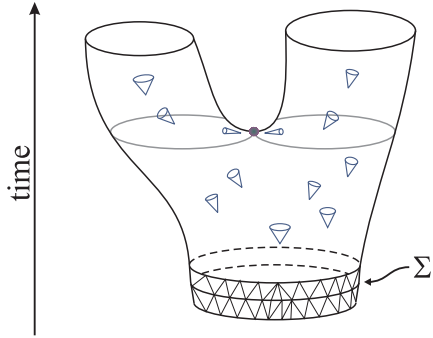}}}}
\caption{The light cone structure (and therefore the underlying Lorentzian geometry) 
becomes degenerate in points where space splits in two.}
\label{figh3}
\end{figure}

Above, our first way of deriving this formulation was as a kind of ``effective" theory:
we started from the set of all
Euclidean two-dimensional geometries of a fixed topology. These geometries
are ``isotropic" in the sense that they do no carry any a priori preferred direction.
We then superimposed
a notion of proper time on them and integrated out part of the degrees
of freedom.
However, when starting in the physically correct {\it Lorentzian} signature,
one can formulate a general principle which excludes geometries
whose {\it spatial} topology is not constant in time \cite{teitelboim}. 
The point is that spatial topology changes are associated with causality 
violations of one kind or other. This is illustrated by the ``trouser geometry"
depicted in Fig.\ \ref{figh3}. As is clear from the embedding of this two-dimensional
spacetime in flat Minkowski
space, with time pointing upward, there must be at least one point near the crotch
of the trousers where the tangent plane is exactly horizontal and the light cone
therefore degenerate.
Note that imposing causality conditions on the geometry to eliminate such
configurations only makes sense
in the presence of a Lorentzian metric and cannot even be formulated in a purely
Euclidean theory, in the absence of any extra structure. 

By the same token, one can take as domain of the path integral the set of all Lorentzian
piecewise flat triangulations whose causal structure is well defined, and where
in particular no changes of spatial topology are allowed to occur.
The set of {\it causal dynamical triangulations} (CDT) -- which can be defined
in any dimension (not just $d=2$) -- obeys a strong version
of causality of this kind, which is implemented by requiring each triangulation
to be the product of a one-dimensional ``triangulation" (a line with equidistant points),
representing discrete proper time, and other triangulated degrees of freedom, representing the
spatial directions of the geometry, which may be thought of as triangulated fibres over
a one-dimensional base space.\footnote{Product triangulations, of which this is
a particular instance, were investigated in \cite{DittrichLoll}, see also \cite{Guitteretal}.}
As an added bonus, each triangulation in the class of CDT  can be
analytically continued to Euclidean signature, and the associated gravitational Regge actions 
satisfy the standard relation between actions defined in spacetimes of 
Lorentzian and Euclidean signatures, namely,
\beq\label{h16}
iS_{Lorentzian} \mapsto -S_{Euclidean}.
\eeq 
Despite the fact that the actions obey \rf{h16},
the Lorentzian theory defined on CDT geometries will even after this ``Wick rotation" 
be {\it distinct} from the full Euclidean theory, because not every Euclidean 
triangulation is the image of a causal, Lorentzian one. The subclass of
Euclidean geometries that are in the image can be obtained ``surgically" as
explained above, by superimposing a notion of proper time on each
Euclidean triangulation and then removing all of its baby universes associated
with spatial topology changes. 
The two-dimensional case is sufficiently simple to allow us to 
perform the calculation in either way, by starting from a path integral over all
Euclidean geometries and removing baby universes, or by starting from a path
integral over causal (CDT) geometries and rotating it to Euclidean signature.
Both results agree after a redefinition of the coupling 
constants. Let us note in passing that our formulation -- not only in dimension 2, but 
also in higher dimensions -- has a couple of characteristics reminiscent
of so-called Ho\v rava-Lifshitz gravity \index{Ho\v rava-Lifshitz Gravity}, 
namely, the use of a preferred time foliation
and a unitary time evolution. We will return to this subject in Sec.\ 
\ref{hlgravity} below.

\subsection{CDT in higher dimensions}

It is not known whether the above-described procedure
of integrating out baby universes in $d=2$ can be generalized to higher 
dimensions in a simple and useful way. 
It implies that at this stage 
we have two a priori unrelated lattice gravity theories in dimension $d>2$, 
one purely Euclidean based on DT and 
one Lorentzian based on CDT. The latter starts out in physical, 
Lorentzian signature, and imposes local causality conditions (nondegeneracy
of local light cones) and a proper-time time foliation.\footnote{Note that there is
no strict physical requirement that individual path integral histories {\it must}
be causal; individual histories are not physical, observable quantities,
only expectation values computed in the ensemble of histories are.}
For calculational purposes, these lattice configurations are then
rotated to Euclidean signature and the path integral over this 
class can in principle be performed. Of course, since the physics one hopes to describe
ultimately by these theories has Lorentzian character, one will have to perform
an ``inverse Wick rotation" back to Lorentzian spacetime eventually, never mind 
whether the computation at an intermediate step took place in a purely Euclidean
or in a Euclideanized Lorentzian framework.

The simplest implementation of Euclidean DT based on the lattice Regge version of 
the Einstein-Hilbert action (the inclusion of a cosmological term being understood) 
does not seem to lead to a theory with an interesting continuum limit.  
Even if this is the case, it is in principle possible that by adding more terms to the
bare lattice action and suitably tuning the associated new coupling constants, 
an interesting continuum theory may emerge after all. This possibility has been
investigated in the past \cite{DTmeasure}, as well as more recently \cite{laiho},
but there is no conclusive evidence at this point that these modified Euclidean models
can reproduce the physical properties of quantum gravity from CDT, the Lorentzian
lattice gravity theory to which we will turn next (see also \cite{cdtreviews} for a variety of 
reviews of the subject). 

\begin{figure}[t]
\centerline{\scalebox{0.5}{\rotatebox{0}{\includegraphics{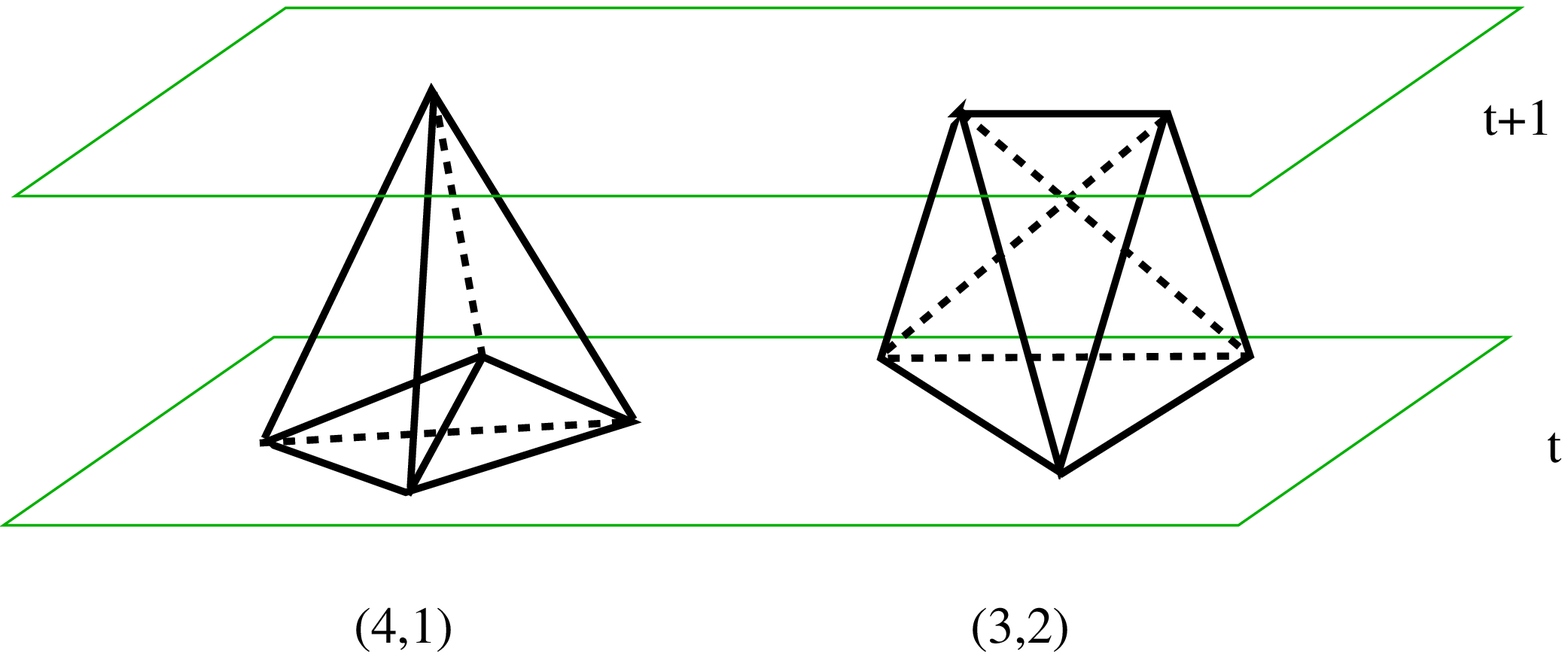}}}
~~~~{\scalebox{0.5}{\rotatebox{0}{\includegraphics{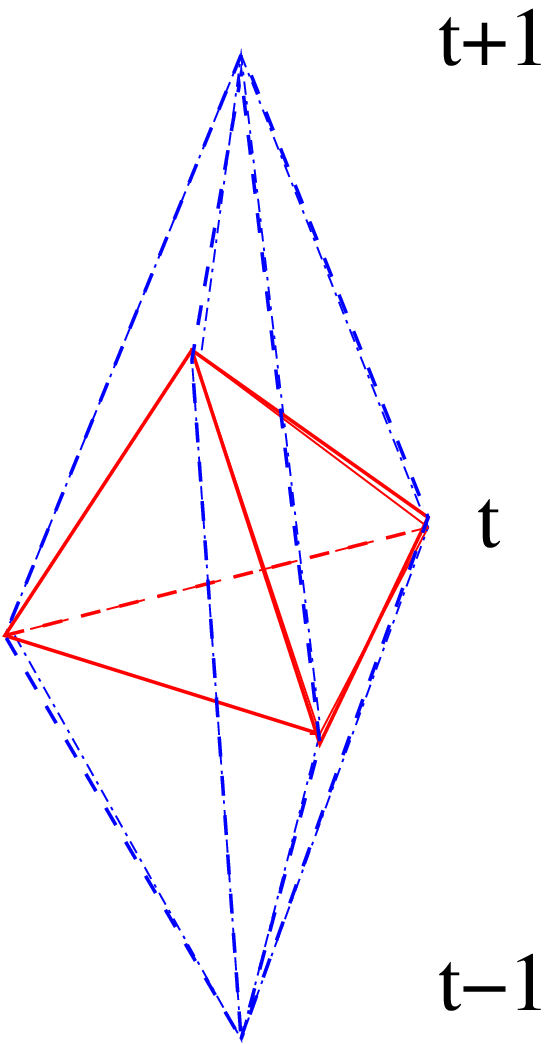}}}}}
\caption{A triangulation in CDT consists of four-dimensional triangulated
layers assembled from (4,1)- and (3,2)-simplices, interpolating between
adjacent integer constant-time slices (left), which in turn are triangulations 
of $S^3$ in terms of equilateral
tetrahedra. Each purely spatial tetrahedron at time $t$ forms the interface between two
(4,1)-simplices, one in the interval $[t-1,1]$, and the other in $[t,t+1]$, as
illustrated on the right. Although a (3,2)-simplex shares none of the five tetrahedra on its
surface with a constant-time slice (the tetrahedra are all Lorentzian), it is nevertheless
needed in addition to the (4,1)-building block to obtain 
simplicial manifolds with a well-defined causal structure.}
\label{figh4}
\end{figure}

Fig. \ref{figh4} illustrates the general construction of a four-dimensional 
CDT triangu\-lation. We take space to be compact and with the 
simplest topology, that of the three-sphere $S^3$. In addition, 
we assume a discrete proper-time foliation and represent the spatial geometry
at each integer proper time $t$ by a three-dimensional simplicial manifold,
given as some configuration of Euclidean DT in terms of equilateral tetrahedra.
By assumption, the tetrahedra are flat in the interior, which means that their geometric
properties are uniquely specified by the length of their edges, which is some number
$a_s>0$ (the same for all edges). To obtain a four-dimensional Lorentzian simplicial
manifold with signature ($-+++$), we still must fill in all intervals $[t,t+1]$ 
between consecutive spatial slices. 
This can be done by using two types of geometrically distinct four-simplices, which again
by assumption are flat in the interior, but this time with Lorentzian signature.
The two different types are the (4,1)- and the (3,2)-simplex
depicted in Fig.\ \ref{figh4}, together with their time-reversed counterparts. The
(4,1)-simplex has as its ``base" one of the spatial tetrahedra contained in 
the triangulated constant-time slice. (The ``4" in the label (4,1) refers to the four
vertices contained in slice $t$ that span this tetrahedron; similarly, the ``1" 
refers to the single vertex shared with slice $t+1$. An analogous labeling has
been used for the (3,2)-simplex.) All that remains to be done to fix the geometry
of the four-simplices is to assign lengths to the edges that have their end points
in adjacent slices, and whose time labels therefore differ by one unit. 
We choose them to be all time-like and of equal (absolute) length
$a_t>0$, which in our signature convention implies that their squared edge length 
is given by $-a_t^2$.\footnote{Note that $a_t$ gives us an approximate distance measure between
adjacent spatial slices labelled by integer-$t$, where the distance of a point in slice
$t+1$ to slice $t$ is defined as the length of the longest geodesic from the point
to the slice.}

Our choice of causal geometries and length assignments has the added benefit that
we can define a map that uniquely maps each Lorentzian CDT history to a Euclidean
DT history. Let us start by parametrizing the relative length of the two lattice parameters
$a_s$ and $a_t$ by a positive real number $\alpha$ defined by $\alpha := -a_t^2/a_s^2$.
Performing a rotation $\a \to -\a$ in the complex lower-half plane can be interpreted as changing
all time-like length assignments of lattice links to space-like ones according to 
\beq\label{h17a}
a_t^2 = -\a a_s^2 ~~~\to~~~ a_t^2 = \a a_s^2.
\eeq  
In order that the Euclidean four-simplices obtained after this rotation satisfy triangle inequalities 
we require $\a > 7/12$. The resulting triangulation represents 
a piecewise linear manifold with {\it Euclidean} signature. If one writes the Lorentzian Regge action
as a function of a single lattice parameter $a:= a_s$ and of $\alpha$,
the action behaves under the rotation \rf{h17a} as one 
would expect na\"ively from a rotation from Lorentzian to Euclidean spacetime, namely,  
\beq\label{h17b}
iS_L[\a] =-S_E[-\a].
\eeq  
The prescription \rf{h17a} leading to \rf{h17b} is the ``Wick rotation" we had in mind 
in our earlier discussion in Sec.\ \ref{sectimeslice}.
It transforms the original Lorentzian path integral with complex weights ${\rm e}^{iS_L(T)}$
to one with real weights ${\rm e}^{-S_E(T)}$, where by slight abuse of notation we use
the same symbol $T$ to denote the initial triangulation (with Lorentzian edge length assignments)
and the one after rotation (which has identical connectivity, but purely Euclidean edge length
assignments). Modulo the sign flip for the length assignments, the domain of the 
Euclideanized path integral is the same set ${\cal T}=\{ T\}$ of triangulations as that of the
original Lorentzian path integral. The set $\cal T$ is of course smaller than the set of 
{\it all} Euclidean triangulations one would obtain by gluing together the same Euclideanized
building blocks, because it still carries an imprint of the causality conditions imposed on
the Lorentzian triangulations.

The fact that in DT and CDT we use {\it standardized} building blocks 
\index{CDT Building Blocks} to 
construct the triangulations means that the Regge action takes on a 
very simple functional form. 
For the special case $|\a|=1$ we have after the Wick rotation
only a single type of building block, the equilateral 
four-simplex with all link lengths equal to $a\equiv a_s$. The Regge form of
the Einstein-Hilbert action becomes
\beq\label{h18}
S_E[-\a =-1; T] = -\kp_0 N_0(T) + \kp_4N_4(T),
\eeq
as is well-known from Euclidean DT quantum gravity. In \rf{h18},
$N_0(T)$ denotes the number of vertices in the triangulation $T$, and
$N_4(T)$ the number of its four-simplices. The coupling $\kappa_0$ is
related to the gravitational coupling constant $G$ via $1/\kp_0 \propto G a^2$,
and $\kp_4$ should be identified with $a^4\La/G$, where 
$\Lambda$ is the cosmological constant.

Whenever $|\a|\not= 1$, we retain the two different building blocks (of type (4,1) and
(3,2)) after the rotation, and the action will depend on their total numbers,
$N_4^{(4,1)}$ and $N_4^{(3,2)}$, separately instead of only on their sum
$N_4=N_4^{(4,1)} \! +\!N_4^{(3,2)}$.
It is convenient to parametrize the resulting Euclideanized Regge action in the form
\bea\label{h19}
S_E[-\a;T] &=& -(\kp_0+6 \Del)N_0(T) +
\kp_4\Big(N_4^{(3,2)}(T)+N_4^{(4,1)}(T)\Big) +\\
&&\Del \Big(N_4^{(3,2)}(T)+2N_4^{(4,1)}(T)\Big),
\nonumber
\eea 
where the asymmetry parameter $\Del$ \index{Asymmetry Parameter $\Del$}
is a function of $\a$ such that 
$\Del(\a\!=\! 1) =0$.

We note that $\Delta$ appears in \rf{h19} on a par with the other 
two coupling constants, $\kp_0$ and $\kp_4$. In what follows, we will treat it
as a third independent coupling constant.
The reason for doing this -- despite the fact that it has no immediate 
interpretation in the Einstein-Hilbert action -- 
is that in the region of phase space (the space spanned by the three couplings 
$\kp_0$, $\kp_4$ and $\Del$)
where we observe interesting, apparently continuum physics, the  
entropy of geometries is as important as the contributions coming from the
bare action term. To make this more explicit, one can rewrite the Euclidean 
partition function of the theory as a sum over the counting variables
$N_4^{(4,1)}$, $N_4^{(3,2)}$ and $N_0$ according to
\bea\label{h20}
Z(\kp_0,\kp_4,\Del) &=& 
\sum_T \e^{-S_E[T]} \\
&=& \sum_{N_4^{(4,1)},N_4^{(3,2)},N_0} 
\e^{-S_E[N_4^{(4,1)},N_4^{(3,2)},N_0]} \cN(N_4^{(4,1)},N_4^{(3,2)},N_0),
\nonumber 
\eea
where $\cN(N_4^{(4,1)},N_4^{(3,2)},N_0)$ is the number of triangulations
with $N_4^{(4,1)}$ four-simplices of type (4,1), $N_4^{(3,2)}$ four-simplices
of type (3,2) and $N_0$ vertices. Introducing the notation 
$c_1 = N_0/N_4^{(4,1)}$ and 
$c_2 = N_4^{(3,2)}/N_4^{(4,1)}$, the leading-order behaviour of this combinatorial quantity 
in the large-volume limit is known to be of the form
\beq\label{h21}
\cN(N_4^{(4,1)},N_4^{(3,2)},N_0) = \e^{f(c_1,c_2)N_4^{(4,1)}+{\rm s.l.}},   
\eeq
where ``s.l.'' denotes subleading terms in $N_4^{(4,1)}$, and the $c_i$ typically have
some boundedness properties.
Since in the same limit the action \rf{h19} can be similarly approximated by 
$S_E= \tilde f(c_1,c_2)N_4^{(4,1)}+{\rm s.l.}$,
it implies that in the region of phase space
where the four-volume can become large, both
$\cN$ and  $e^{-S_E}$ have the same functional form and are potentially of the same magnitude.
It turns out that this is the same region where we observe interesting continuum-like 
physics. Because of contributions from both ``energy" and ``entropy", it is  
clear therefore that the effective action governing physics in this non-perturbative
region can be very
different from the ``na\"ive'' Einstein-Hilbert action, justifying our inclusion of
$\Delta$ as a tunable parameter in the bare action.

To summarize: taking as our starting point spacetimes with Lorentzian signature,
we can consider the transition amplitude between an initial and a final  
spatial three-geometry, $[g^{(3)}_{\rm i}]$ and $[g^{(3)}_{\rm f}]$
separated by a proper time $t$.
We can then regularize the theory, using CDT, representing three-geometries by
equilateral Euclidean triangulations and spacetime geometries by causal, Lorentzian
triangulations with a discrete proper-time foliation. In the CDT framework,
each of the latter can be rotated to Euclidean signature,
leading to a regularized, Euclideanized sum-over-histories. What remains to be done
is to ``remove the regulator", that is, take the lattice spacing $a$ to zero. 
Denoting the initial and final spatial triangulations by $T^{(3)}_i$ and $T^{(3)}_f$,
we thus arrive at the prescription 
\beq\label{h22}  
G_E([g^{(3)}_{\rm i}],[g^{(3)}_{\rm i}] ,t,\kp_0,\kp_4,\Del) :=
 \lim_{a\to 0}  \;\;\; \sum_{T:T^{(3)}_i \to T^{(3)}_f}
 \e^{-S_E[T]},
\eeq
which can be viewed as the four-dimensional 
generalization of the two-dimensional loop-loop amplitude 
$\tG(\tilde{\ell}_1,\tilde{\ell}_2,t)$ introduced in \rf{h14} above.
For a more detailed description of the CDT construction we 
refer the interested reader to \cite{3d4d,cdtlecture,physrep}.

\section{The phase diagram}

Contrary to the situation in two dimensions, we cannot calculate 
the amplitude \rf{h22} analytically. 
However, we can extract a lot of non-trivial, non-perturbative information by performing
Monte Carlo computer simulations \index{Monte Carlo Simulations}. 
This will usually start with an investigation of the structure of the space of coupling constants (the ``phase
space" of the underlying statistical system), in particular, trying to identify regions associated with
a second-order phase transition, where according to standard lore one can hope 
to obtain continuum physics \index{Second Order Phase Transition}. 

Let us highlight two technical aspects related to our implementation of the 
computer simulations. Firstly, rather 
than fixing specific boundary three-geometries $T^{(3)}$ at times 0 and $t$,
we take time to be periodic. Although this is strictly speaking in contradiction with
imposing causality (it introduces closed time-like curves), in practice it turns out to not
affect results. The nature of the ground states of geometry is such that by
choosing $t$ sufficiently 
large -- assumed from now on -- the boundary condition becomes irrelevant. 

Secondly, as we have discussed, the action \rf{h19} depends on
three coupling constants, one of which, $\kp_4$, can 
be identified with the cosmological coupling constant, 
multiplying the spacetime volume $V$ in the action.
In the computer simulations it is convenient to keep this four-volume 
fixed, which means that the cosmological constant does not really play a role.
We compensate for this by performing 
separate simulations at different (fixed) spacetime volumes.
From these we can in principle reconstruct results which depend on the 
cosmological constant via a Laplace transformation,
\beq\label{h23}
G(\kp_4,....) = \int_0^\infty dV  \; \e^{\kp_4 V} \, G(V,....).
\eeq
We are therefore left with two coupling constants, $\kp_0$ and $\Del$.
\begin{figure}[t]
\centerline{\scalebox{0.8}{\rotatebox{0}{\includegraphics{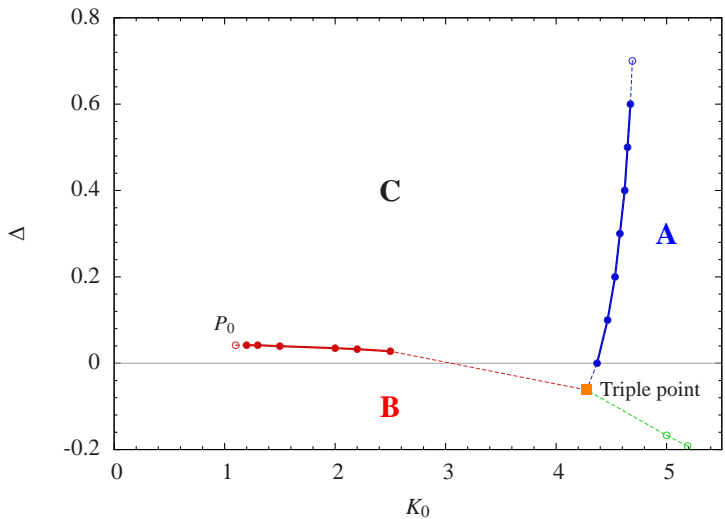}}}}
\caption{The phase diagram of CDT quantum gravity in the $(\kp_0,\Del)$-plane.}
\label{figh5}
\end{figure}
The corresponding phase diagram is shown in Fig.\ \ref{figh5} \cite{phasediagram}
and exhibits three distinct phases, labelled A, B and C. 
Phase C appears to be the one relevant for continuum physics, because only there 
do we observe extended four-dimensional universes \cite{ABC}. 
A careful numerical analysis reveals strong evidence that the transition between 
phases C and A is first order, whereas between 
phases C and B we find a second-order transition \cite{jordan}. 
This very exciting result implies that the B-C phase transition
line is a candidate for a region in the coupling-constant 
plane where genuine UV continuum limits may exist, defined by 
approaching specific points on the line. Conversely, moving away
from the transition line into phase C corresponds to going 
towards an IR limit.  

\subsection{Phase C}

\begin{figure}[t]
\centerline{\scalebox{0.7}{\rotatebox{0}{\includegraphics{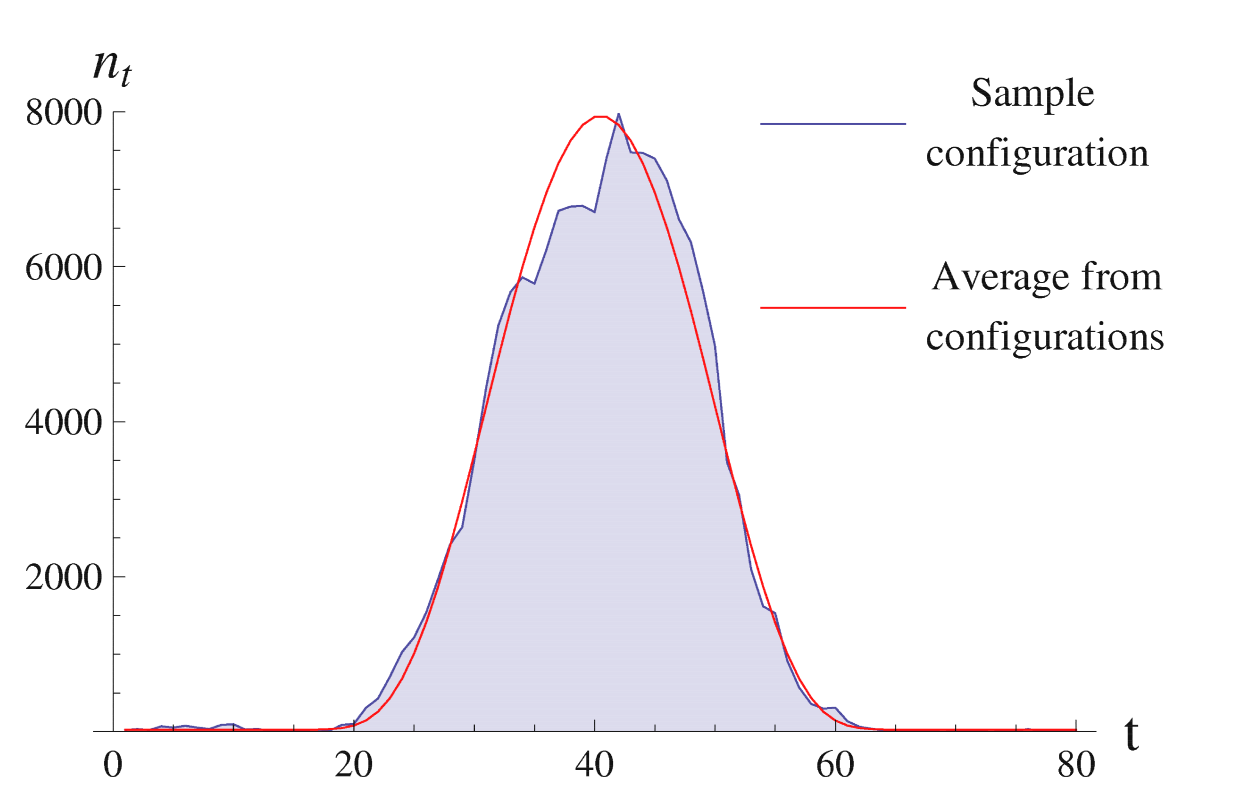}}}}
\caption{The three-volume of spatial slices as a function of proper time in phase
C. Shown are a sample configuration of the volume profile, as well as the
expectation value of the same quantity.}
\label{figh6}
\end{figure}

The reason why phase $C$ is related to extended four-dimensional spacetimes
is illustrated in Fig.\ \ref{figh6}, which shows both a 
sample path-integral configuration generated 
by the computer during the Monte Carlo simulations, 
as well as the associated quantum observable, obtained by averaging in
the ensemble. 
While of course we have access to the complete geometric 
information of the quantum spacetimes that are generated, only a single
degree of freedom is depicted here, the three-volume of a spatial
slice of the quantum spacetime as a function of proper time. 
The time extension in a given simulation is always fixed (in the case at hand
to 80 discrete time steps). What we observe in Fig.\ \ref{figh6} is that
the universe does not make use of the full time interval
available, but has a non-vanishing volume only on a connected subset of
the time axis.\footnote{Since we impose the kinematical constraint that
the spatial volume at fixed $t$ cannot become smaller than 5 tetrahedra --
the minimal number required to build a simplicial manifold of topology $S^3$ --
the volume never vanishes completely. More precisely, what we observe in addition to the
bell-shaped part of the volume profile is the formation of
a distinct ``stalk" which is close to the minimal size of 5 everywhere.} 

A quantitative piece of evidence in favour of a {\it four-}dimensional
extended universe is the fact that its time extension (not counting the stalk)
scales like $N_4^{1/4}$ when the total discrete four-volume $N_4$ of the universe 
used in the simulations is varied. Similarly, its discrete three-volume $N_3(t)$ scales
like $N_4^{3/4}$. 
Contrary to one's na\"ive expectations, these findings are highly non-trivial, because 
they have been derived in a non-perturbative, background-independent path integral
formulation. The simplicial building blocks of our regularization {\it are} four-dimensional,
but since assembling them is only dictated by the Boltzmann weight $e^{-S_E[T]}$ {\it without}
any reference to a four-dimensional background, there is no reason why the
resulting object, extrapolated to infinite lattice volume, should be four-dimensional on any scale.

This is specifically true in the non-perturbative regions of phase space where the 
entropic contributions to the effective action compete with those coming from the classical bare 
action, as explained above. In these regions it can easily happen
that a type of configuration is entropically favoured that has no resemblance at all with
an extended four-dimensional universe. Just from looking at the volume profiles,
it is obvious that something like this does indeed 
happen in phases A and B, which as a result do not appear to have any classical limit
resembling general relativity \cite{ABC}. However, even in phase C the observed quantum
universe is truly an outcome of non-perturbative dynamics, not a consequence of the dominance 
of the classical action.\footnote{Since we are working in Euclidean signature, dominance of
the classical action would be fatal for the path integral, because of the action's unboundedness
from below. In phase C, this instability is cured by the entropy of ``microstates" or, in other words,
the path-integral measure \cite{semi,measure}.} 

The fact that the path-integral measure \index{Path Integral Measure}
can play a crucial role 
in determining the non-perturbative dynamics was a main lesson learned 
already earlier in the context of four-dimensional DT quantum gravity. When one 
considers a path integral ensemble of geometries obtained from gluing
four-dimensional equilateral Euclidean simplices, with the only constraint 
that the topology should be that of $S^4$, one ends 
up with a universe of vanishing linear
extension and infinite Hausdorff dimension \cite{4dDT}. 
This makes the situation depicted in Fig.\ \ref{figh6} all the more remarkable!

\subsection{The effective action}

However, the surprises do not stop here. The smooth curve in Fig.\ \ref{figh6} represents
the expectation value of the volume profile, that is, the average over path integral 
configurations measured in the Monte Carlo simulations. For $N_4$ sufficiently large 
this curve is very precisely fitted by the function 
\beq\label{h24}
 \la N_3(i)\ra  \propto 
N_4^{{3}/{4}}
\cos^3 \left(\frac{i}{s_0 N_4^{1/4}}\right),
\eeq
where $i$ denotes (integer) lattice time, $N_4$ the total 
number of four-simplices and $N_3(i)$ the number of tetrahedra 
at time $i$ \cite{s4a,s4b}, and $s_0$ is a constant.\footnote{The formula is of course not valid in the stalk, where
$N_3(i) \approx 5$.}   

Can the functional form of the expectation value found in \rf{h24} be obtained 
directly from an action principle? The answer is yes \cite{semi}.
A long time ago, Hartle and Hawking explored a minisuperspace approach 
to quantum gravity \index{Minisuperspace}, 
where all gravitational (field) degrees of freedom at a fixed time
are represented by a single number, the so-called scale factor or, equivalently,
the total three-volume of the universe.\footnote{This rather crude approximation is 
borrowed from standard cosmology, where 
homogeneity and isotropy are assumed to give a realistic description of our universe on the 
very largest scales.}
Taking this classically reduced formulation as the starting point of the quantization, 
finding a quantum theory of gravity is reduced to a quantum mechanical 
problem in one variable, the scale factor $a(t)$ \cite{hh}. 

The volume profile \rf{h24} of the emergent extended universe found in phase C of CDT quantum
gravity can be derived from an ``effective" action \index{Effective Action}
for the three-volume, namely,
\beq\label{h25}
S_{eff}= \frac{1}{24\pi G} \int d t 
\left( \frac{ \dot{V_3}^2(t)}{V_3(t)}+k_2 V_3^{1/3}(t)
-\lam V_3(t)\right),
\eeq
where $t$ denotes proper time, $k_2$ is a numerical constant
and $\lam$ is a Lagrange multiplier, not a cosmological constant, 
because the total four-volume $V_4$ is kept fixed in the simulations. 
Intriguingly, one obtains exactly the same expression (up to an overall sign) when
plugging a spatially homogeneous and isotropic ansatz for the metric $g_{\mu\nu}(x)$ into
the Euclidean Einstein-Hilbert action, and re-expressing the dependence on the scale factor 
in terms of the three-volume $V_3(t)\propto a^3(t)$.
The solution to the equations of motion derived from \rf{h25} is the Euclidean
de Sitter universe (a round four-sphere), which
as a function of proper time $t$ results in the 
$\cos^3(t/V_4^{1/4})$-dependence of eq.\ \rf{h24}. 

Despite the fact that they lead to very similar results for the dynamics of the
scale factor, let us stress that 
conceptually there is a big difference between the ansatz of 
Hartle and Hawking, who simply assumed a minisuperspace reduction from the
outset, and studying the effective dynamics of (the expectation value of) the scale factor in a
full theory of quantum gravity, as we are doing. 
The only small but important reminder of the non-perturbative origin of the action
\rf{h25} is its overall sign, which is opposite to that found in Euclidean cosmology.
It can be attributed directly to ``entropic'' contributions to the effective action.
The solutions to the equations of motion are of course not affected by this sign difference.
A discretization of the effective action \rf{h25} has the functional form 
\beq\label{h26}
S_{discr} =
k_1 \sum_i \left(\frac{(N_3(i+1)-N_3(i))^2}{N_3(i)}+
\tilde{k}_2 N_3^{1/3}(i)-\tilde{\lam} N_3(i)\right).
\eeq
We have managed to reconstruct it in detail from the simulation
data for the volume-volume correlator $\la V_3(t)V_3(t')\ra$, and have also shown that  
the quantum fluctuations around the de Sitter ``background geometry'' are
well described by the action \rf{h26}, yet another non-trivial result \cite{s4b}.

The same data have allowed us to relate the continuum coupling constant 
$G$ in \rf{h25} to the constant $k_1$ in \rf{h26} according to
\beq\label{h27}
G = \frac{a^2}{k_1} \frac{\sqrt{C_4}\; s_0^2}{3\sqrt{6} },
\eeq
where $a$ is the lattice spacing and $C_4$ is essentially
the volume of a four-simplex (for lattice spacing $a\!=\! 1$), but depends weakly
on the ratio between $N_4^{(1,4)}$ and $N_4^{(2,3)}$ 
(since the (4,1)- and (3,2)-simplices only have identical four-volumes when $\a\!=\! 1$).
This ratio, as well as the value of  the constant $s_0$, defined in eq.\ \rf{h24}, 
depend on the choice of the bare coupling constants $\kp_0$ and $\Del$ in phase C.

Let us consider a typical choice for these couplings, $(\kp_0,\Del)=(2.2,0.6)$,
positioning us in the interior of phase C. At this point in phase space, we have 
measured $k_1$ 
and with the help of \rf{h27} expressed Newton's constant and the Planck length
$\ell_P$ in terms of the lattice spacing, resulting in 
\beq\label{h28}
G\approx 0.23 a^2,~~~~~ \ell_{P} \equiv \sqrt{G} \approx 0.48 a.
\eeq
From the identification of spacetime with a Euclidean de Sitter universe
we have that $V_4 =  8\pi^2 R^4/{3} = C_4 N_4 a^4$, where $C_4$ is the same 
quantity that appeared in \rf{h27}.
For the range of four-volumes used in the simulations, $N_4\in[45.000,360.000]$,
the linear size $\pi R$ of the quantum de Sitter universes 
\index{Quantum de Sitter Universe} lies between 12 and 21
Planck lengths $\ell_{P}$. The small size of our universes is compatible 
with the fact that the observed quantum fluctuations in the three-volume are quite
substantial, as illustrated by Fig.\ \ref{figh6} (see also Fig.\ \ref{figh7}). 
For larger universes, the volume fluctuations
will quickly become irrelevant.
 
However, in order to investigate quantum properties
of spacetime at Planckian and even sub-Planckian length scales, we want to do the
opposite, namely, make the universes smaller and in this way increase the small-scale resolution of the 
simulations. How can we
improve on \rf{h28} such that a single Planck length $\ell_P$ corresponds not to just half a lattice 
spacing, but to many lattice spacings $a$? From eqs.\ \rf{h27} and \rf{h28} it is clear
that when $k_1$ goes to zero, $\ell_P$ can become much larger than $a$. 
The question is whether we can adjust $k_1$ to go to zero. 
Since $k_1$ depends on the bare coupling constants $\kp_0$ and $\Del$,
we have performed a scan of phase C to determine its qualitative behaviour \cite{s4b}. 
Moving toward the A-C phase transition, $k_1$ is indeed decreasing, without
going all the way to zero in the range of coupling constants scanned so far. 
Approaching the B-C phase transition is more difficult, because  
the system undergoes a second-order transition, and we observe a 
corresponding critical slowing-down. 
As far as we can tell from the numerical data at this stage, $k_1$ does {\it not} 
decrease when we
approach this transition. However, as we will see in the next section, having $k_1$ go to
a fixed value different from zero is actually the behaviour predicted at an ultraviolet 
second-order transition line, and therefore compatible with the continuum scenario
we have appealed to earlier.

\subsection{Making contact with asymptotic safety}

Let us return to the renormalization group equation\ \rf{h1}, which was formulated 
in terms of the dimensionless coupling constant $\tG = G\,E^2$. 
Now that we have a UV cut-off, the lattice link length $a$, we can 
instead form the dimensionless quantity $\hG = G/a^2$. From 
\rf{h27} it can essentially be identified with the inverse of $k_1$,
which we can measure. We can reformulate the renormalization 
group in terms of the new short-distance cut-off as
\beq\label{h29} 
 G(a) = a^2 \hG(a),~~~~ a \frac{\d \hG}{\d a} = -\b(\hG),~~~~
\b(\hG) = 2\hG -c \hG^2 +\cdots,
\eeq
where $c$ depends on the constant $\omega$ of eq.\ \rf{h1}.
Near the putative non-Gaussian UV fixed point $\hG^*$, we can expand $\hG$
and $k_1$ to lowest order in $a$ according to
\beq\label{h30a}
\hG(a) = \hG^* - K a^\tc,~~~k_1(a) = k_1^* + \tilde{K} a^\tc,
\eeq
for some $K$, $\tilde K$, where the approach to the fixed point is governed by
the exponent
\beq\label{h30b}
\tc=-\b'(\hG^*).
\eeq
As explained in Sec.\ \ref{2.1}, in standard lattice theory one would now relate
the lattice spacing near the fixed point to the bare coupling constants 
with the help of some correlation length $\xi$. However, in four-dimensional
quantum gravity we do not yet have a suitable correlation length at our disposal
which could play this role.

In search of an alternative, let us first consider the equation $V_4 = N_4 a^4$, which defines 
the dimensionful continuum four-volume
$V_4$ in terms of the number $N_4$ of four-simplices and the lattice
spacing. If we could 
consider $V_4$ as fixed, we could replace the $a$-dependence of \rf{h30a} by a
$N_4$-dependence, with the advantage that $N_4$ is a parameter we can straightforwardly control.
Re-expressing eq.\ \rf{h30a} in terms of $N_4$ yields
\beq\label{h31}
k_1(N_4) = k_1^* -K' N_4^{-\tc /4},
\eeq
for some $K'$. Since we can measure $k_1$, we could determine the flow to the fixed point.
The question is now which lattice measurements we should perform in order to make eq.\ \rf{h31}
applicable. Increasing $N_4$ while {\it staying}
at a specific point $(\kp_0,\Del)$ in phase C does {\it not} correspond to keeping $V_4$ fixed,  
because during this process the size of the quantum fluctuations in the three-volume decreases 
relative to the expectation value of the three-volume. (More precisely,
we already know that the ratio goes to zero like $1/N_4^{1/4}$.) 
Conversely, if ``physics" is to be constant,
which includes a constant $V_4$, that same ratio should also remain constant.

We will use this observation as our definition for what we mean by a ``path of constant physics". 
If we had a correlation length available, we could increase 
$N_4$ and simultaneously {\it change} the bare coupling constants in such a way 
that the ratio of the correlation length to the linear extension of 
the universe of volume $N_4$ (both in terms of lattice units) stayed constant. 
\begin{figure}[t]
\centerline{\scalebox{0.6}{\rotatebox{0}{\includegraphics{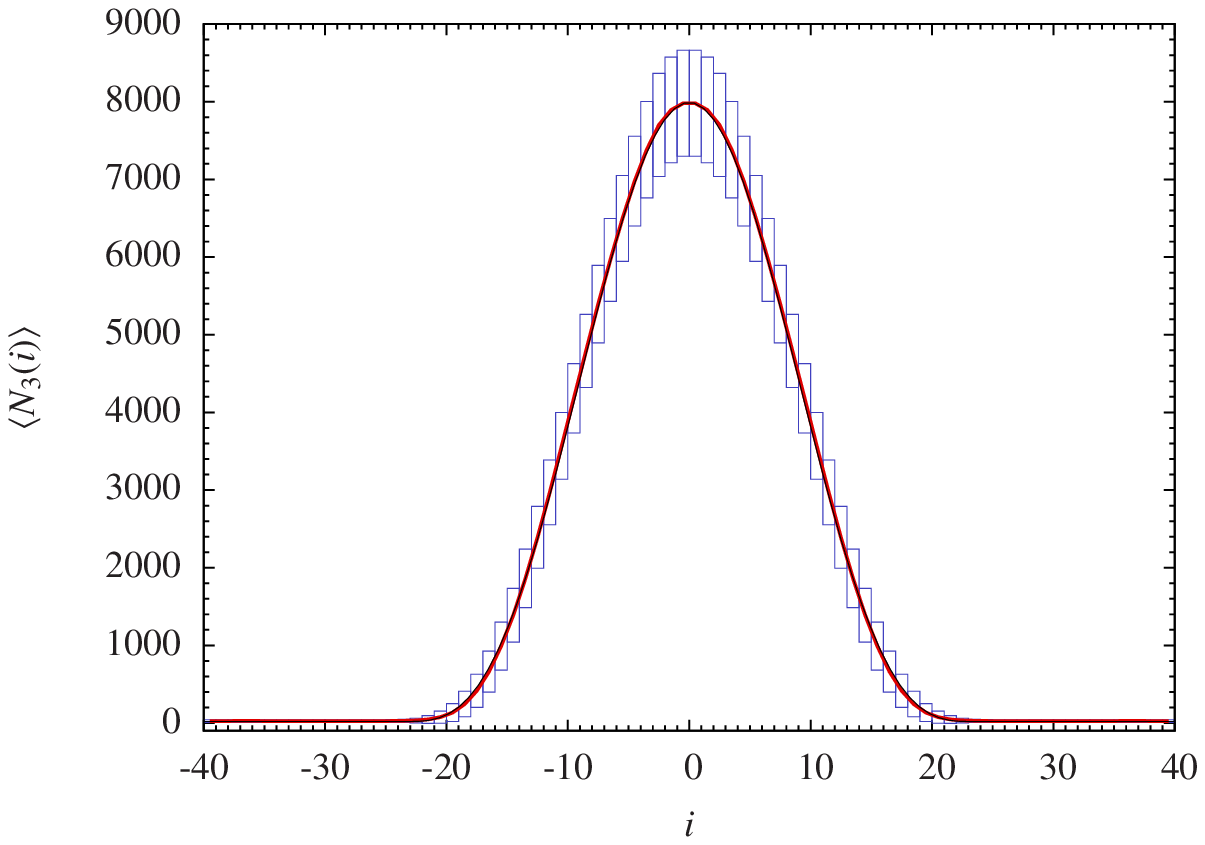}}}
\scalebox{0.3}{\rotatebox{0}{\includegraphics{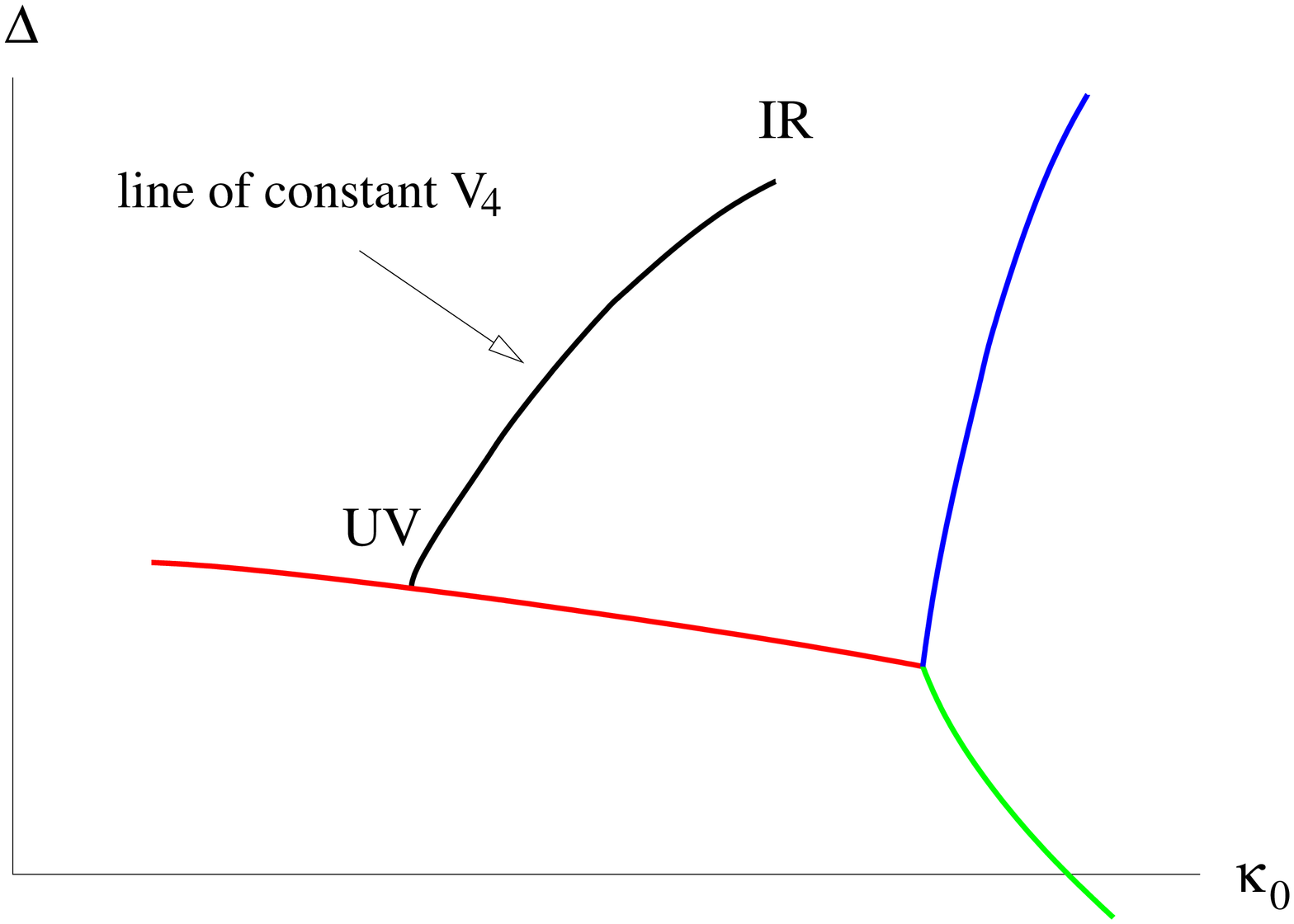}}}}
\caption{Left: Three-volume profile for given $N_4$, for specific values 
$(\kp_0,\Del)$ of the bare coupling constants. Also indicated is the 
magnitude of the three-volume fluctuations around the mean value. 
While the expectation value of the three-volume scales like $N_4^{3/4}$, the fluctuations only scale like
$N_4^{1/2}$. Right: identifying a path of ``constant physics" in the $\kappa_0$-$\Delta$ plane.
Starting at some point in phase C, a path moving toward the UV phase transition is created 
by increasing $N_4$ and simultaneously adjusting $\kappa_0$ and $\Delta$, such that 
the ratio of the size of the three-volume fluctuations and the expectation value of the three-volume
remains constant.}
\label{figh7}
\end{figure}
In the absence of a suitable correlation length, we will use the magnitude of
the three-volume fluctuations instead, and identify a ``path of constant physics"
as a trajectory in phase C along which the discrete four-volume $N_4$ grows, but
the accompanying change in the bare couplings $\kappa_0$ and $\Delta$ ensures that
the three-volume fluctuations likewise increase, in such a way that the ratio between the magnitude of
the fluctuations and the mean three-volume stays the same.
Fixing this ratio forces us to change 
bare coupling constants when we increase $N_4$, in this way tracing out 
a path that moves toward one of the phase transitions 
bordering phase C, see Fig.\ \ref{figh7} (right) 
for a schematic illustration. 
Preliminary results from computer simulations to determine the flow defined in this way
indicate that it should start quite close to the B-C phase transition if it should resemble
the flow line of constant physics shown in the figure, raising again the issue
of critical slowing-down near the B-C line.

\section{Relation to Ho\v rava-Lifshitz gravity}\label{hlgravity}

As described above, our CDT data in phase C can be fitted well to the 
functional form \rf{h26},
which in turn can be seen as a discretized version of the minisuperspace 
action \rf{h25}. 
There is a residual ambiguity in the interpretation of 
the discrete time coordinate appearing 
in the identification \rf{h24}, which can be thought of as an overall,
finite scaling between the time and spatial directions.
As we have emphasized, due to the entropic nature of the effective action,
there is no compelling reason to take the geometric
length assignments of the regularized theory literally. 
We have identified the time ``coordinate'' $t$ with continuum proper
time in such a way that we obtain a round four-sphere, which  
is a perfectly legitimate and physically well-motivated choice.
However, as we vary the bare couplings $\kp_0$ and $\Del$, 
the overall shape of the computer-generated
universe changes in terms of the number of
lattice spacings in the time direction relative to
those in the spatial directions.
Although this change is qualitatively in agreement with
the change of $\a$ as a function of $\kp_0$ and $\Del$, there
is no detailed quantitative agreement.

Instead of choosing continuum time to be consistent with a continuum $S^4$-geometry
as one moves in phase space, one may be able to find a
modified action which describes the observed behaviour without
performing an overall time rescaling which depends on $\kappa_0$ and $\Del$. 
This may be especially appropriate in the vicinity of the phase
transition, where the length scales one is probing become increasingly Planckian,
and one would expect significant contributions to the effective dynamics from
terms not contained in the infrared form of the Einstein-Hilbert action including
higher-order curvature terms.

We will consider yet another generalization, which suggests itself because of the built-in
anisotropy between time and space of the CDT set-up, namely, a deformation \`a la 
Ho\v rava-Lifshitz \cite{horava}. 
A corresponding effective Euclidean continuum action, 
including measure contributions, and expressed
in terms of standard metric variables could be of the form
\beq\label{horava}
S_H = \frac{1}{16\pi G} \int \d^3x\ \d t \; N \sqrt{g}
\Big((K_{ij}K^{ij}-\lam K^2) + (-\g R^{(3)} +2\La + V(g_{ij})\Big),
\eeq
where $K_{ij}$ denotes the extrinsic curvature and $g_{ij}$ the three-metric
of the spatial slices, $R^{(3)}$ the corresponding three-dimensional scalar
curvature, $N$ the lapse function, and finally $V(g_{ij})$ a ``potential''
which in Ho\v rava's continuum
formulation would contain higher orders of spatial derivatives, potentially
rendering $S_H$ renormalizable. In our case we are not committed
to any particular choice of potential $V(g_{ij})$, since we are
not imposing renormalizability of the theory in any conventional
sense. 

An effective $V(g_{ij})$ could be generated
by entropy, i.e.\ by the measure, and may not relate
to any discussion of the theory being renormalizable.
The kinetic term depending on the extrinsic
curvature is the most general such term which is at most second order in
time derivatives and consistent with spatial diffeomorphism invariance.
The parameter $\lambda$ appears in the (generalized) DeWitt metric, which
defines an ultralocal metric on the classical space of all
three-metrics\footnote{The value of $\lambda$ governs the signature of
the generalized DeWitt metric
$$
G_\lambda^{ijkl}=\frac{1}{2}\sqrt{\det g} (g^{ik} g^{jl}+g^{il}g^{jk}-
2\lambda g^{ij} g^{kl}),
$$ 
which is positive definite for $\lambda <1/3$, indefinite
for $\lambda =1/3$ and negative definite for $\lambda >1/3$. 
The role of $\lambda$
in three-dimensional CDT quantum gravity has been analyzed 
in detail in \cite{cdtlambda}.}, and
the parameter $\gamma$ can be related to a relative scaling between
time and spatial directions.
Setting $\lam =\g =1$ and $V=0$ in \rf{horava} we recover the standard (Euclidean)
Einstein-Hilbert action.


Making a simple minisuperspace ansatz with compact spherical slices,
which assumes homogeneity and isotropy of the spatial three-metric $g_{ij}$,
and fixing the lapse to $N=1$, the Euclidean action (\ref{horava}) becomes
a function of the scale factor $a(t)$
(see also \cite{elias,brandenberger,calcagni}, as well as \cite{bh} for related work in 2+1
dimensions), that is,
\beql{mini}
S_{mini} = \frac{2 \pi^2}{16\pi G} \int \d t \; a(t)^3 \Big( 3(1-3\lambda)\
\frac{\dot{a}^2}{a^2} -\gamma\ \frac{6}{a^2} +2 \Lambda+ \tV(a)\Big).
\eeq
The first three terms in the parentheses define the IR limit (which in Ho\v rava-Lifshitz gravity is 
assumed to include a flowing of $\lambda$ to its ``GR value"), 
while the potential term 
$\tV(a)$ contains inverse powers of the scale factor $a$ coming from possible
higher-order spatial derivative terms.

Our reconstruction of the effective action from the computer data
is compatible with the functional form \rf{mini} of the minisuperspace action.
If we were able to extract the constant $\tilde{k}_2$ in front of the potential term in
\rf{h26}, it would enable us to fix the ratio $(1-3\lam)/2\g$ appearing in \rf{mini} \cite{semiclassical}.
At this stage, the precision of our measurements is insufficient to do so. 
The same is true for our attempts to determine $\tV(a)$ for small values 
of the scale factor, which is important for understanding UV quantum corrections
to the potential near $a(t)=0$. 
Once we have developed a better computer algorithm which allows us to 
approach the B-C phase transition line more closely, investigating such Planckian  
properties and testing scenarios of Ho\v rava-Lifshitz type will be within reach.

\subsection{Conclusions}

In constructing a theory of quantum gravity using Causal Dynamical Triangulations, 
one of our initial inputs was the Regge action,
which appears in the weights of individual spacetimes in the gravitational path integral.
However, as we have emphasized repeatedly, the full effective action generated dynamically
by performing the non-perturbative
sum over histories is only indirectly related
to this ``bare'' action. Likewise, the coupling constant $k_1$, which
appears in front of the effective action and we view as related to the
gravitational coupling constant $G$, has no obvious direct
relation to the ``bare'' coupling $\kp_0$ appearing in the Regge action.

Nevertheless, the leading terms in the
effective action for the scale factor are
precisely the ones present in \rf{h25} or, more generally,
in the effective Ho\v rava-Lifshitz action \rf{mini}, at least for sufficiently large
values of the scale factor.
The fact that a kinetic term quadratic in derivatives appears as the
leading term in the effective action is perhaps less surprising,
but that the correct powers of the (undifferentiated) variable $N_3(i)$ appear 
in both the kinetic and potential terms in \rf{h26} is rather remarkable
and very encouraging for the entire CDT quantization program. 

For the range of bare coupling constants and four-volumes investigated until
now our results are compatible with the Einstein-Hilbert action.
Better data and more observables 
will be required to discriminate between a ``pure gravity" behaviour and
an anisotropic deformation \`a la Ho\v rava-Lifshitz in the deep ultraviolet.
A beautiful feature of CDT quantum gravity is that entirely
non-perturbative questions of this kind can be formulated explicitly and 
addressed with the non-perturbative lattice tools available, and -- if one is
lucky -- be answered quantitatively.

\subsection*{Acknowledgment}

JA and AG thank the Danish Research Council for financial
support via the grant "Quantum gravity and the role of black holes", 
and the EU for support through the ERC Advanced
Grant 291092, "Exploring the Quantum Universe" (EQU).
JJ acknowledges a partial support of the International PhD Projects Programme of the Foundation
for Polish Science within the European Regional Development Fund of
the European Union, agreement no. MPD/2009/6. 
 RL acknowledges support through several
Projectruimte grants by the Dutch Foundation for Fundamental Research on Matter (FOM).


\end{document}